\documentclass[pre,twocolumn]{revtex4}
\usepackage[english]{babel}
\usepackage{amsmath}
\usepackage{amssymb}
\usepackage{epsfig}

\begin{document}
\title{Nonuniformly Filled Vortex Rings in Nonlinear Optics}
\author{Victor P. Ruban}
\email{ruban@itp.ac.ru}
\affiliation{Landau Institute for Theoretical Physics RAS,
Chernogolovka, Moscow region, 142432 Russia}

\date{\today}

\begin{abstract}
A new type of long-lived solitary structures for paraxial
optics with two circular polarizations of light in a
homogeneous defocusing Kerr medium with an anomalous
group velocity dispersion has been revealed numerically
in the coupled nonlinear Schrödinger equations.
A found hybrid three-dimensional soliton is a vortex 
ring against the background of a plane wave in one
of the components, and the core of the vortex is filled
with another component nonuniformly in azimuth angle. 
The existence of such quasistationary structures with 
a reduced symmetry in a certain parametric region is
due to the saturation of the so-called sausage
instability caused by the effective surface tension of
a domain wall between two polarizations.

\vspace{1mm}

\noindent V. P. Ruban, JETP Lett. {\bf 117}(8), 583 (2023); 
DOI: 10.1134/S0021364023600817
\end{abstract}

\maketitle

\subsection*{Introduction}

As known, the character of the propagation of a
weakly nonlinear quasimonochromatic light wave in a
three-dimensional optical medium with the Kerr
nonlinearity is determined both by the type of group
velocity dispersion (normal or anomalous) and by the
type of nonlinearity (focusing or defocusing). 
Individual coherent structures and individual dynamical
regimes, which are approximately described by the
nonlinear Schr\"odinger equation (NSE) with corresponding 
coefficients, are characteristic of each of the four 
possible combinations (see, e.g., [1-4] and references 
therein). In particular, a wave in a defocusing medium 
with anomalous dispersion is similar to a stable 
quantum liquid (dilute Bose-Einstein condensate
of mutually repulsing cold atoms; the NSE in this case
is usually called the Gross-Pitaevskii equation). Such
a system allows not only dark solitons but also
topological excitations in the form of quantized vortices
[5]. An analogy is even closer if the optical wave carries
two circular polarizations because light in this case is
described by two coupled NSEs [6] as well as the
binary Bose-Einstein condensate [7-13]. However,
some difference in the problems exists because real
Bose-Einstein condensates are spatially confined by
the confining potential of a trap, whereas a model
without the external potential is more realistic for
light in a homogeneous medium.

It is remarkable that the existence of the second
component makes it possible to significantly extend
the set of accessible coherent structures. In particular,
domain walls separating regions with the right and left
circular polarizations in addition to dark solitons and
quantized vortices in each of two components are possible 
in the phase separation regime typical of nonlinear 
optics [14-21]. A domain wall is characterized by
an effective surface tension [10, 22], which affects the
equilibrium shape and dynamics of domains. Various
aspects of phase separation in Bose-Einstein condensates 
were considered in [23-41]. It is noteworthy that
combined vortex–soliton complexes exist [9, 19-21, 42-48],
in particular, a vortex ring in one of the components 
whose core forms a potential well for the other
component and is filled by it [49]. In this work, new
results are obtained for such structures in application
to nonlinear optics.

\subsection*{Preliminary remarks}

For trapped binary Bose-Einstein condensates, it was 
recently shown theoretically [49] that a filled ring
in an external quadratic axisymmetric potential of the
trap can be both linearly stable and unstable with
respect to azimuthal perturbations, depending on
parameters. Instability appearing at an increase in the
amount of the trapped second component is similar to
the so-called sausage instability of a hollow vortex in
classical hydrodynamics. Its nature can be easily
understood considering a straight vortex with axisymmetric 
perturbations. When the width of the domain wall is 
much smaller than the local radius of the core
of the filled quantum vortex, the classical two-liquid
model allows one to write the functional of the effective 
potential energy whose minimum determines the
stable distribution of the second component along the
axis of the vortex (see [47] and references therein). 
The homogeneous configuration becomes unstable when
the amount of the filling component per unit length
exceeds a certain critical value. In this case, the minimum 
of the functional is reached on sausage-shaped
structures (aspherical bubbles). An accurate analytical
description in the case of the filled ring is very difficult, 
but instability is qualitatively similar. The specificity 
of the system in the trap is that this instability is
not saturated. The unstable filled vortex ring in the
condensate is finally destroyed after a complex interaction
(accompanied by energy transfer) between
inhomogeneous flows of the second component along
the core and motions of the ring itself against the
inhomogeneous density background [49].

In this work, it is shown that the situation in a
homogeneous space is different. The filled optical vortex 
ring against the homogeneous wave background
holds its integrity and propagates at long distances,
even being subjected to sausage instability. The difference
compared to the trapped Bose-Einstein condensate 
is the absence of a noticeable energy transfer from
the degrees of freedom determining the spatial position
of the ring to the degrees of freedom describing
the azimuthal distribution of the second component.
The behavior of the system (for the ring with a given
radius) depends on the intensity of the wave and the
amount of the second component. At a small amount
of the second component or at a relatively low power
of the background wave, the initial azimuthal inhomogeneity
of filling leads to subsequent alternating thickening 
of the ring on two opposite sides. Such a behavior 
is characteristic of a one-dimensional nonlinear
oscillator with an even single-well potential (in a rough
approximation, the amplitude of the first azimuthal
harmonic of the distribution of the second component
along the ring can be considered as the coordinate of this
oscillator). At a large filling or at a strong nonlinearity, 
the thickening remains at its place and a significant
transfer of the second component along the ring
does not occur. Such a configuration corresponds to
the symmetry breaking in the system with a one-dimensional
double-well potential and indicates the
saturation of the sausage instability.

It is also noteworthy that an analog of the vortex
ring in the two-dimensional space is a vortex pair. Two
vortices unequally filled with the second component
(see [43] and references therein) vaguely resemble the
considered situation. A rougher analog is an asymmetric 
pair of so-called dark-bright solitons in one
dimension (see [50] and references therein). One of
the main features of the three-dimensional ring is the
possibility of longitudinal flows along the vortex,
whereas ``the bright'' component in the one-dimensional 
and two-dimensional cases can flow from one
potential well (formed by the vortex or ``dark''
component) to the other only through tunneling.

\subsection*{Model}

Let a three-dimensional optically transparent
homogeneous dielectric medium have an isotropic
dispersion relation for linear waves 
$k(\omega)=\sqrt{\varepsilon(\omega)}\omega/c$,
an anomalous group velocity dispersion (i.e., $k''(\omega)<0$
in a certain frequency range), and defocusing 
Kerr nonlinearity. The anomalous dispersion
range is usually near the low-frequency edge of the
transparency window (usually the infrared range in
real materials; see, e.g., [51, 52]).

The displacement nonlinear in electric field at the
frequency $\omega$ is assumed in the form
\begin{equation}
{\bf D}^{(3)}=\alpha(\omega)|{\bf E}|^2{\bf E}
+\beta(\omega)({\bf E}\cdot{\bf E}) {\bf E}^*,
\label{D3}
\end{equation}
where $\alpha(\omega< 0)$ and $\beta(\omega)< 0$. Defining the amplitudes
$A_{1,2}$ of the left and right circular polarizations by the formula
\begin{equation}
{\bf E}=\big[({\bf e}_x+i{\bf e}_y) A_1 
           + ({\bf e}_x-i{\bf e}_y) A_2 \big]/\sqrt{2}
\end{equation}
and substituting Eq. (1) into Maxwell’s equations, one
can derive the standard system of the coupled NSEs
for $A_{1,2}$ (see [6, 14-21]).

Let the carrier frequency $\omega$, the corresponding
wavenumber $k_0=2\pi/\lambda_0$, and the first, $k_0'$, and 
second, $k_0''$, derivatives be fixed. As accepted in optics, the
distance $\zeta$ along the axis of the beam serves as the evolution 
variable instead of the time $t$, and the ``delayed'' time
$\tau=t-\zeta/v_{\rm gr}$, where $1/v_{\rm gr}=k_0'$, serves as 
the third ``spatial'' coordinate. Let the large radius of the vortex
ring $R_0\sim 10^2 \lambda_0$ be the scale for the transverse 
coordinates. The longitudinal coordinate $\zeta$, the variable $\tau$,
and the electric field are represented in units of
$k_0 R_0^2\sim 10^5\lambda_0$, $R_0\sqrt{k_0 |k_0''|}$,  and 
$\sqrt{2\epsilon/|\alpha|}/(k_0 R_0)$, respectively. 
In terms of these dimensionless variables, the system of the 
coupled NSEs for slowly varying complex envelopes 
$A_{1,2}(x,y,\tau,\zeta)$ has the form
\begin{equation}
i\frac{\partial A_{1,2}}{\partial \zeta}=\Big[-\frac{1}{2}\Delta 
+|A_{1,2}|^2+ g_{12}|A_{2,1}|^2\Big]A_{1,2},
\label{A_12_eqs}
\end{equation}
where $\Delta=\partial_x^2+\partial_y^2+\partial_\tau^2$  is the
three-dimensional Laplace operator in the  $(x,y,\tau)$ ``coordinate''
space. The cross phase modulation parameter $g_{12}=1+2\beta/\alpha$
in the typical case is approximately 2. The condition
$g_{12} > 1$ corresponds to the phase separation mode.

\subsection*{Numerical method}

The numerical simulation was performed using the
standard split-step Fourier method of the second
order of accuracy in the evolution variable $\zeta$. An
important difference from previous works [21, 40, 41]
is the method of preparation of the initial state. Since
the perfect vortex ring is not static [in the $(x,y,\tau)$
space] and moves steadily at a certain ``velocity'' $u$
(along the $\tau$) axis for definiteness), it is convenient
to make the change
\begin{eqnarray}
A_1(x,y,\tau,\zeta)&=&\sqrt{I_0}
A(x,y,\tau-u\zeta,\zeta)e^{-iI_0\zeta},\\
A_2(x,y,\tau,\zeta)&=&\sqrt{I_0}
B(x,y,\tau-u\zeta,\zeta)e^{(iu\tau-i\mu_2\zeta)},
\end{eqnarray}
where $I_0$ is the intensity of the unperturbed plane wave and 
$\mu_2$ is the ``chemical potential'' of the second component. 
In the strictly steady-state case, new unknown functions 
$A(x,y,z,\zeta)$ and $B(x,y,z,\zeta)$ would be independent of
$\zeta$, and $A\to 1$ far from the ring (the new ``spatial'' 
variable $z=\tau-u\zeta$ differs from the evolution variable $\zeta$).

The new variables seemingly give no advantage for
the simulation of a conservative system for $A$ and $B$
that follows from Eqs.(3). However, the aim of this
change of variables is to use the imaginary-time propagation 
method to numerically prepare the initial state
as close as possible to a steady state. It should be
emphasized that the stationary solution of interest is a
saddle point of the corresponding functional
${\cal H}_u[A,B]-\int(I_0 |A|^2+\mu_2|B|^2) dx dy dz$
rather than its strict minimum. In particular, this functional
as a function of the radius of the vortex ring has a maximum; 
correspondingly, the size of the ring deviates slowly 
from the initial value in the dissipative procedure 
while hard modes relax. The pseudotime interval
in the imaginary-time propagation method has to be
limited; as a result, perturbations in the initial state
cannot be removed completely. At the same time, the
simulation with unsteady perturbations makes the
numerical experiment more realistic and allows one to
generally estimate the stability of studied structures.

All calculations were carried out in the $(2\pi)^3$ cube
with periodic boundary conditions in the variables $x$,
$y$, and $z$. Consequently, the presented structures are
strictly speaking not completely isolated. However,
the effect of periodically located ``neighboring'' identical 
vortices is not too significant if the radius of the ring is 1.

\begin{figure}
\begin{center}
\epsfig{file=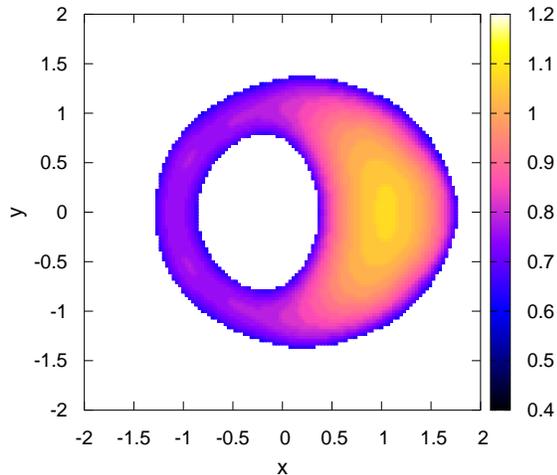, width=84mm}
\end{center}
\caption{(Color online) Surface $|A(x,y,z)|^2=0.5$ of the
numerically obtained inhomogeneously filled quasistationary 
vortex ring at the propagation distance $\zeta=100$, where the $z$ 
coordinate is given in color (details see in the main text).
}
\label{surface} 
\end{figure}

\begin{figure}
\begin{center}
(a)\epsfig{file=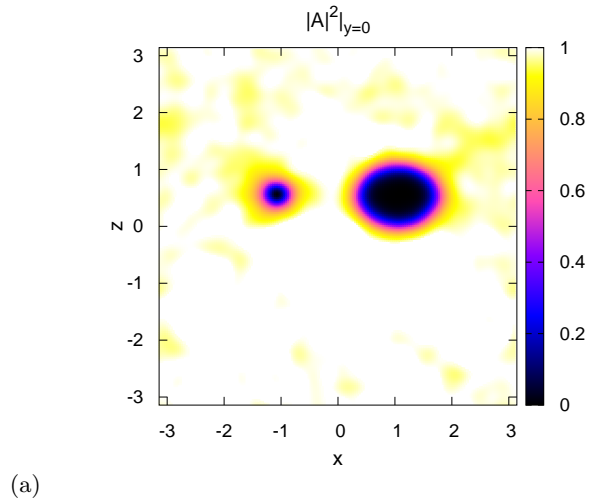, width=79mm}\\
(b)\epsfig{file=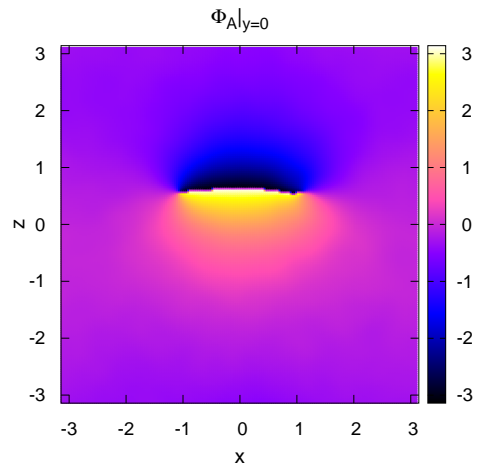, width=79mm}\\
(c)\epsfig{file=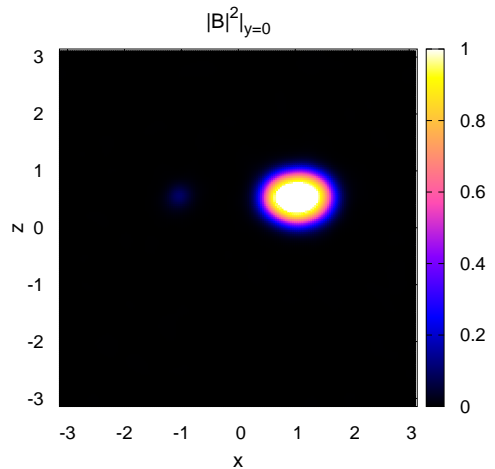, width=79mm}
\end{center}
\caption{
(Color online) (a) Relative intensity and (b) the
phase of the first component, as well as (c) the relative
intensity of the second component in the section of the
vortex ring shown in Fig.1 by the $y=0$ plane.
}
\label{sections} 
\end{figure}

\subsection*{Typical parameters}

Selecting the parameters $u$, $I_0$, and $\mu_2\sim I_0$, as well
as the bare functions $A$ and $B$ and the pseudotime
interval, for the imaginary-time propagation procedure, 
one can ensure a fairly ``good'' initial state with a
desired amount of the second component $N_2=\int |B|^2 dx dy dz$.

Since the dimensionless width of the core of the unfilled 
vortex can be estimated as $\xi\sim 1/\sqrt{I_0}$, of the
most interest are the values $I_0=16$--$80$, at which the
dimensional width of the core is much smaller than
the radius of the ring, but is still much larger than the
wavelength for the applicability of the quasimonochromatic 
approximation. The $\tau$ width of the core should satisfy the
constraint $\xi R_0\sqrt{k_0 |k_0''|}\omega\gtrsim 2\pi$, which
is more stringent than $\xi k_0 R_0\gtrsim 2\pi$ in the case
$(\omega^2|k_0''|/k_0)<1$.

The parameter $u$ should be selected self-consistently 
such that the subsequent motion of the ring in the $(x,y,z)$ 
space occurs at the minimum possible velocity (in practice, 
it is sufficient to ensure the residual velocity $\lesssim 0.2$ ).
The $u$ values in this case are in the range from
$\approx 1.5$ at $I_0=16$ to $\approx 2.0$  at $I_0=81$.

The evolution interval in the variable $\zeta$ was no
shorter than 60. In all cases, the vortex-soliton complex 
did not tend to a noticeable increase in amplitudes 
of perturbations, which indicates that the considered 
structures are long-lived and practically stable.

\begin{figure}
\begin{center}
\epsfig{file=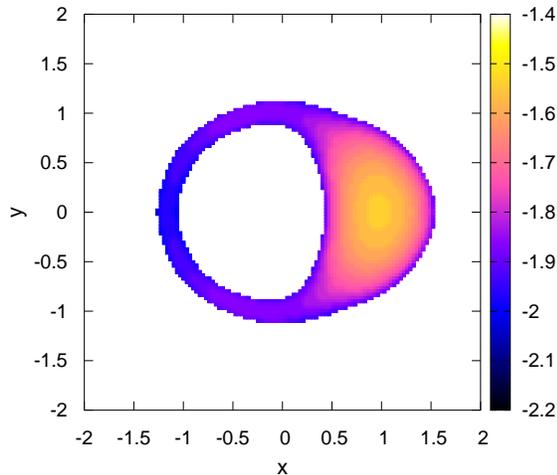, width=84mm}
\end{center}
\caption{
Numerical example of the saturation of the strong sausage 
instability at $I_0=81$, $u\approx 2.0$, $N_2\approx 1.0$.
}
\label{surface-2} 
\end{figure}

\subsection*{Examples}

Figure 1 presents the first example of the filled
quasi-steady vortex ring with broken azimuthal symmetry 
obtained in the numerical experiment with the parameters 
$I_0=25$, $u\approx 1.65$, and $N_2\approx 1.8$. 
The corresponding profiles of the intensities $|A|^2$ and $|B|^2$,
as well as the vortex phase, are shown in Fig.2. For comparison, 
the sausage instability in the case of a smaller amount 
of the second component $N_2\approx 0.6$ was absent
because the amplitude of the first azimuthal mode
oscillated passing through zero (without slowing near zero). 
The critical filling value appeared $N_{2,c}\approx 1.0$. At
the low intensity $I_0=16$, the sausage instability did not
occur even for the strongly filled (``thick'') vortex with
$N_2\approx 1.9$. On the contrary, at the high intensity $I_0=36$,
the critical filling value decreased to $N_{2,c}\approx 0.5$.

Since the dimensionless effective surface tension coefficient is 
$\sigma\propto \sqrt{I_0}$, the most pronounced sausage
instability should be expected at high intensities $I_0$. An
example of the strongly inhomogeneous vortex ring
obtained at the parameters $I_0=81$, $u\approx 2.0$, and
$N_2\approx 1.0$ is presented in Fig.3. It is seen that the
second component is almost completely absent in a significant
part of the ring; this component is displaced to a clearly
separated bubble, so that the configuration has the shape of
a bag with a handle or a kettlebell.

\subsection*{Conclusions}

To summarize, it has been numerically demonstrated 
that the filled vortex ring in the defocusing
optical Kerr medium with the anomalous group velocity 
dispersion at a sufficiently high intensity of the
background wave shows the sausage instability. However, 
unlike similar objects in trapped binary Bose-Einstein 
condensates of cold atoms, instability in this case is 
saturated and results in previously unknown long-lived 
fundamentally three-dimensional nonaxisymmetric
structures. Their characteristic size is estimated 
at about 100 $\mu$m and they propagate to meters.

It is also noteworthy that the parameter $I_0$ in the
presence of a relatively weak linear absorption of light
decreases exponentially during the propagation of the
wave. If the symmetry of the initial quasistationary
ring is broken, the structure at a certain distance $\zeta$
should approach a regime without the sausage instability 
and restore the axial symmetry on average. The corresponding
numerical experiments (their results are not presented in
this work) have confirmed this scenario.

\subsection*{Funding}

This work was supported by the Ministry of Science and
Higher Education of the Russian Federation (state assignment 
no. 0029-2021-0003).

\subsection*{Conflict of interest}

The author declares that he has no conflicts of interest.

\end{document}